\documentclass[aps,prl,twocolumn,showpacs,groupedaddress]{revtex4}
\usepackage{color}
\usepackage[dvips]{graphicx}
\usepackage{amssymb}
\newcommand{\T}{{\cal T}}
\newcommand{\R}{{\cal R}}
\newcommand{\V}{{\cal V}}

\begin{document}
%\preprint{}
\title{Finite bias visibility of the electronic Mach-Zehnder interferometer}
\author{Preden Roulleau}
\author{F. Portier}
\author{D. C. Glattli}
\altaffiliation[Also at ]{LPA, Ecole Normale Sup\'erieure, Paris.}
\author{P. Roche}
\email{patrice.roche@cea.fr}
\affiliation{Nanoelectronic group, Service de Physique de l'Etat Condens\'e,\\
CEA Saclay, F-91191 Gif-Sur-Yvette, France}
\author{A. Cavanna}
\author{G. Faini}
\author{U. Gennser}
\author{D. Mailly}
\affiliation{CNRS, Phynano team, Laboratoire de Photonique et Nanostructures,\\
Route de Nozay, F-91460 Marcoussis, France}
\date{\today}
\begin{abstract}
We present an original statistical method to measure the
visibility of interferences in an electronic Mach-Zehnder
interferometer in the presence of low frequency fluctuations. The
visibility presents a single side lobe structure shown to result
from a gaussian phase averaging whose variance is quadratic with
the bias. To reinforce our approach and validate our statistical
method, the same experiment is also realized with a stable sample.
It exhibits the same visibility behavior as the fluctuating one,
indicating the intrinsic character of finite bias phase averaging.
In both samples, the dilution of the impinging current reduces the
variance of the gaussian distribution.
\end{abstract}

 \maketitle

Nowadays quantum conductors can be used to perform experiments
usually done in optics, where electron beams replace photon beams.
A beamlike electron motion can be obtained in the Integer Quantum
Hall Effect (IQHE) regime using a high mobility two dimensional
electron gas in a high magnetic field at low temperature. In the
IQHE regime, one-dimensional gapless excitation modes form, which
correspond to electrons drifting along the edge of the sample. The
number of these so-called edge channels corresponds to the number
of filled Landau levels in the bulk. The chirality of the
excitations yields long collision times between quasi-particles,
making edge states very suitable for quantum interferences
experiments like the electronic Mach-Zehnder interferometer (MZI)
\cite{Ji03Nature422p415,Samuelsson04PRL92n026805,Neder07CM0705.0173}.
Surprisingly, despite some experiments which show that equilibrium
length in chiral wires is rather long \cite{Machida97SSC103p441},
very little is known about the coherence length or the phase
averaging in these "perfect" chiral uni-dimensional wires. In
particular, while in the very first interference MZI experiment
the interference visibility showed a monotonic decrease with
voltage bias, which was attributed to phase noise
\cite{Ji03Nature422p415}, in a more recent paper, a surprising
non-monotonic decrease with a lobe structure was observed
\cite{Neder06PRL96p016804}. A satisfactory explanation has not yet
been found, and the experiment has so far not been reported by
other groups to confirm these results.

We report here on an original method to measure the visibility of
interferences in a MZI, when low frequency phase fluctuations
prevent direct observation of the periodic interference pattern
obtained by changing the magnetic flux through the MZI. We studied
the visibility at finite energy and observed a single side lobe
structure, which can be explained by a gaussian phase averaging
whose variance is proportional to $V^2$, where $V$ is the bias
voltage. To reinforce our result and check if low frequency
fluctuation may be responsible for that behavior, we realized the
same experiment on a stable sample : we also observed a single
side lobe structure which can be fitted with our approach of
gaussian phase averaging. This proves the validity of the results,
which cannot be an artefact due to the low frequency phase
fluctuations in the first sample. In both samples, the dilution of
the impinging current has an unexpected effect : it decreases the
variance of the gaussian distribution.

\begin{figure}[t]
\centerline{\includegraphics[angle=-90,width=8cm,keepaspectratio,clip]{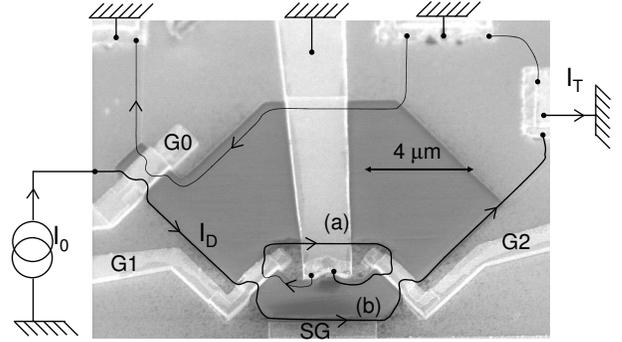}}
\caption{SEM view of the electronic Mach-Zehnder with a schematic
representation of the edge state. G0, G1, G2 are quantum point
contacts which mimic beam splitters. The pairs of split gates
defining a QPC are electrically connected via a Au metallic bridge
deposited on an insolator (SU8). G0 allows a dilution of the
impinging current, G1 and G2 are the two beam splitters of the
Mach-Zehnder interferometer. SG is a side gate which allows a
variation of the length of the lower path (b).}\label{Photo.fig}
\end{figure}

The MZI geometry is patterned using e-beam lithography on a high
mobility two dimensional electron gas in a GaAs/Ga$_{1-x}$Al$_x$As
heterojunction with a sheet density
$n_S=2.0\times10^{11}$~cm$^{-2}$ and a mobility of
$2.5\times10^{6}$~cm$^2$/Vs. The experiment was performed in the
IQHE regime at filling factor $\nu=n_Sh/eB=2$ (magnetic field
$B=$5.2 Tesla). Transport occurs through two edge states with an
extremely large energy redistribution length
\cite{Machida97SSC103p441}. Quantum point contacts (QPC)
controlled by gates G0, G1 and G2 define electronic beam splitters
with transmissions $\T_0$, $\T_1$ and $\T_2$ respectively. In all
the results presented here, the interferences were studied on the
outer edge state schematically drawn as black lines in
Fig.(\ref{Photo.fig}), the inner edge state being fully reflected
by all the QPCs. The interferometer consists of G1, G2 and the
small central ohmic contact in between the two arms. G1 splits the
incident beam into two trajectories (a) and (b), which are
recombined with G2 leading to interferences. The two arms defined
by the mesa are 8~$\mu$m long and enclose a 14~$\mu$m$^2$ area.
The current which is not transmitted through the MZI,
$I_B=I_D-I_T$, is collected to the ground with the small ohmic
contact. An additional gate SG allows a change of the length of
the trajectory (b). The impinging current $I_0$ can be diluted
thanks to the beam splitter G0 whose transmission ${\cal T}_0$
determines the diluted current $dI_D = {\cal T}_{0}\times dI_0$.
We measure the differential transmission through the MZI by
standard lock-in techniques using a 619 Hz frequency
5~$\mu$V$_{rms}$ AC bias $V_{AC}$ superimposed to the DC voltage
$V$. This AC bias modulates the incoming current $dI_D =\T_0\times
h/e^2\times V_{AC}$, and thus the transmitted current in an energy
range close to $eV$, giving the transmission $\T(eV)=dI_T/dI_0$.

Using the single particle approach of the Landauer-B{\"u}ttiker
formalism, the transmission amplitude $t$ through the MZI is the
sum of the two complex transmission amplitudes corresponding to
paths (a) and (b) of the interferometer; $t =t_0\{
t_1\exp(i\phi_{a}) t_2 -r_1\exp(i\phi_{b})r_2\}$. This leads to a
transmission probability $\T(\epsilon) =
\T_0\{\T_1\T_2+\R_1\R_2+\sqrt{\T_1\R_2\R_1\T_2}\sin[\varphi(\epsilon)]\}$,
where $\varphi(\epsilon)=\phi_{a}-\phi_{b}$ and
$\T_i=|t_i|^2=1-\R_i$. $\varphi(\epsilon)$ corresponds to the
total Aharonov-Bohm (AB) flux across the surface $S(\epsilon)$
defined by the arms of the MZI, $\varphi(\epsilon)= 2\pi
S(\epsilon)\times eB/h$. The surface $S$ depends on the energy
$\epsilon$ when there is a finite length difference $\Delta
L=L_a-L_b$ between the two arms. This leads to a variation of the
phase with the energy,
$\varphi(\epsilon+E_F)=\varphi(E_F)+\epsilon\Delta L/(\hbar v_D)$,
where $v_D$ is the drift velocity. When varying the AB flux, the
interferences manifest themselves as oscillations of the
transmission; in practice this is done either by varying the
magnetic field or by varying the surface of the MZI with a side
gate
\cite{Ji03Nature422p415,Neder06PRL96p016804,Litvin07PRB75n033315}.
The visibility of the interferences defined as $\V =
(\T_{MAX}-\T_{MIN})/(\T_{MAX}+\T_{MIN})$, is maximum when both
beam splitter transmission are set to 1/2. In the present
experiment the MZI is designed with equal arm lengths ($\Delta
L=0$) and the visibility is not expected to be sensitive to the
coherence length of the source $\hbar v_D/max(k_B T,eV_{AC})$.
Thus the visibility provides a direct measurement of the
decoherence and/or phase averaging in this quantum circuit.
\begin{figure}[h]
%06_11_14.visibilite.opj[PAPER]
\centerline{\includegraphics[angle=-90,width=9cm,keepaspectratio,clip]{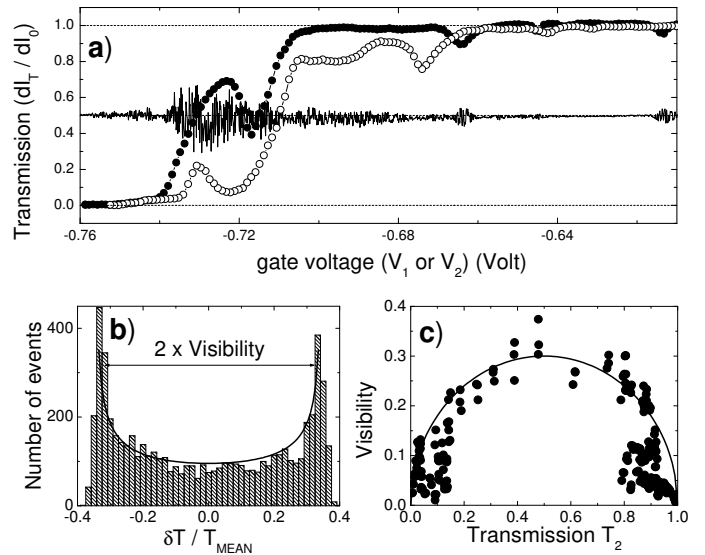}}

\caption{ \textbf{Sample \#1} \textbf{a)}Transmission $\T =
dI_T/dI_0$ as a function of the gate voltages $V1$ and $V2$
applied on G1 and G2. ($\circ$) $\T=\T_1$ versus $V1$. ($\bullet$)
$\T=\T_2$ versus $V2$. The solid line is the transmission $\T$
obtained with $\T_1$ fixed to 1/2 while sweeping $V2$~:
transmission fluctuations due to interferences with low frequency
phase noise appears. \textbf{b)} Stack histogram on 6000
successive transmission measurements as a function of the
normalized deviation from the mean value. The solid line is the
distribution of transmission expected for a uniform distribution
of phases. \textbf{c)}Visibility of interferences as a function of
the transmission $\T_2$ when $\T_1=1/2$. The solid line is the
$\sqrt{\T_2(1-\T_2)}$ dependence predicted by the
theory.}\label{Trans.fig}
\end{figure}

In Ref.\cite{Ji03Nature422p415}, ~60\% visibility was observed at
low temperature, showing that the quantum coherence length can be
at least as large as several micrometer at 20 mK (and probably
larger if phase averaging is the limiting factor). At finite
energy (compared to the Fermi energy), the visibility was also
found decreasing with the bias
voltage\cite{Ji03Nature422p415,Litvin07PRB75n033315,Neder06PRL96p016804}.
This effect is not due to an increase of the coherence length of
the electron source which remains determined by $eV_{AC}$ or $k_B
T$ \cite{Chung05PRB72p125320}. In a first experiment, a monotonic
visibility decrease was found, which was attributed to phase
averaging, as confirmed by shot noise measurements
\cite{Ji03Nature422p415}. Nevertheless, it remains unclear why and
how the phase averaging increases with the bias. In a recent
paper, instead of a monotonic decrease of the visibility, a lobe
structure was observed for filling factor less than 1 in the QPCs
\cite{Neder06PRL96p016804}. No non-interacting electron model was
found to be able to explain this observation, and although
interaction effects have been proposed
\cite{Sukhorukov06CM0609288}, a satisfactory explanation has not
yet been found to account for all the experimental observations.
So far, two experiments have shown to two different behaviors,
raising questions about the universality of these observations.
Here, we report experiments where different samples give
consistent results, with a fit to the data clearly demonstrating
that our MZI suffers from a gaussian phase averaging whose
variance is proportional to $V^2$, leading to the single side lobe
structure of the visibility.

We have used the following procedure to tune the MZI. We first
measure independently the two beam splitters' transparencies
versus their respective gate voltages, the inner edge state being
fully reflected. This is shown in figure
(\ref{Trans.fig}\textbf{a}) where the transmission ($\T_1$ or
$\T_2$) through one QPC is varied while keeping unit transparency
for the other QPC. This provides the characterization of the
transparency of each beam splitter as a function of its gate
voltage. The fact that the transmission vanishes for large
negative voltages means that the small ohmic contact in between
the two arms can absorb all incoming electrons, otherwise the
transmission would tend to a finite value. This is very important
in order to avoid any spurious effect in the interference pattern.
In a second step we fix the transmission $\T_1$ to 1/2 while
sweeping the gate voltage of G2 (solid line of figure
(\ref{Trans.fig}\textbf{a})). Whereas for a fully incoherent
system the $\T$ should be $1/2\times(\R_2+\T_2)=1/2$, we observe
large temporal transmission fluctuations around 1/2. We show in
the following that they result from the interferences, expected in
the coherent regime, but in presence of large low frequency phase
noise. This is revealed by the probability distribution of the
transmissions obtained when making a large number of transmission
measurements for the same gate voltage. Figure
(\ref{Trans.fig}\textbf{b}) shows a histogram of $\T$ when making
6000 measurements (each measurement being separated from the next
by 10~ms). The histogram of the transmission fluctuations
$\delta\T=\T-\T_{mean}$ displays two maxima very well fitted using
a probability distribution
$p(\delta\T/\T_{mean})=1/(2\pi\sqrt{1-(\delta\T/\T_{mean})^2/\V^2})$
(the solid line of figure (\ref{Trans.fig}\textbf{b})). This
distribution is obtained assuming interferences
$\delta\T=\T_{mean}\times \V sin(\varphi)$ and a uniform
probability distribution of $\varphi$ over $[-\pi,+\pi]$. Note
that the peaks around $|\delta\T/\T_{mean}|=\V$ have a finite
width. They correspond to the gaussian distribution associated
with the detection noise which has to be convoluted with the
previous distribution.

\begin{figure}
%06_11_24.opj[PAPER]
%06_11_28.opj[PAPER]
\centerline{\includegraphics[angle=-90,width=9cm,keepaspectratio,clip]{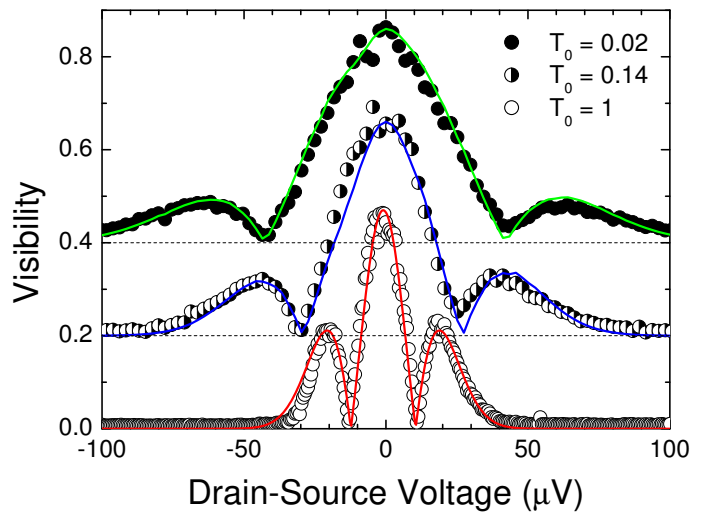}}
\caption{(Color online) \textbf{Sample \#1 :} Visibility of the
interferences as a function of the drain-source voltage $I_0
h/e^2$ for three different values of $\T_0$. The curves are
shifted for clarity. The energy width of the lobe structure is
modified by the dilution whereas the maximum visibility at zero
bias is not modified. Solid lines are fits using equation (1).
From top to bottom, $\T_0=0.02$ and $V_0=31$~$\mu$V, $\T_0=0.14$
and $V_0=22$~$\mu$V, $\T_0=1$ and
$V_0=11.4$~$\mu$V.}\label{Lobe.fig}
\end{figure}
Although no regular oscillations of transmission can be observed
due to phase noise, we can directly extract the visibility of the
interferences by calculating the variance of the fluctuations (the
approach is similar to measurements of Universal Conductance
Fluctuations via the amplitude of 1/f noise in diffusive metallic
wires) \footnote{All the results on the visibility reported here
on sample \#1 have been obtained using the following procedure~:
we measured $N=2000$ times the transmission and calculated the
mean value $\T_{mean}$ and the variance $<\delta\T^2>$. It is
straightforward to show that the visibility is $\V
=\sqrt{2}\sqrt{<\delta\T^2>-<\delta\T^2>_0}/\T_{mean}$, where
$<\delta\T^2>_0$ is the measurement noise which depends on the AC
bias amplitude, the noise of the amplifiers and the time constant
of the lock-in amplifiers (fixed to 10 ms), measured in absence of
the quantum interferences.}. As expected when $\T_1=1/2$, the
visibility extracted by our method is proportional to
$\sqrt{\T_2(1-\T_2)}$, definitively showing that fluctuations
results from interference: we are able to measure the visibility
of fluctuating interferences (see figure (2\textbf{c})).

The visibility depends on the bias voltage with a lobe structure
shown in figure (\ref{Lobe.fig}), confirming the pioneering
observation \cite{Neder06PRL96p016804}. Nevertheless, there are
marked differences. The visibility shape is not the same as that
in ref.\cite{Neder06PRL96p016804}. We have always seen only one
side lobe, although the sensitivity of our measurements would be
high enough to observe a second one if it existed. Moreover, the
lobe width (see figure (\ref{Lobe.fig})) can be increased by
diluting the impinging current with G0, whereas no such effect is
seen for G1 and G2. This apparent increase of the energy scale
cannot be attributed to the addition of a resistance in series
with the MZI because G0 is close to the MZI, at a distance shorter
than the coherence length.
\begin{figure}
%06_11_24.opj[PAPER]
%06_11_28.opj[PAPER]
%\centerline{\includegraphics[angle=-90,width=9cm,keepaspectratio,clip]{Sample2Tuning.eps}}
\centerline{\includegraphics[angle=-90,width=8cm,keepaspectratio,clip]{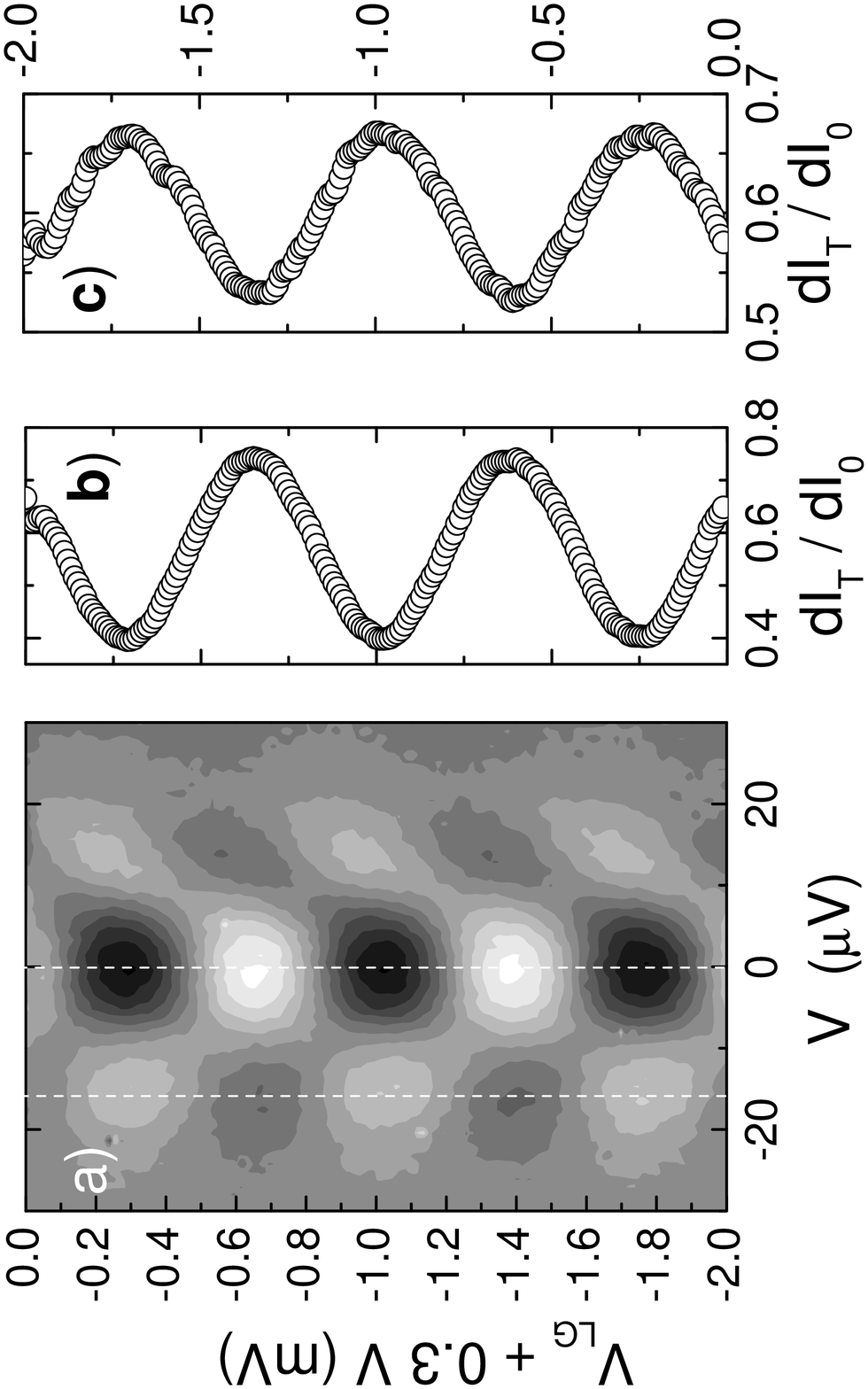}}
\centerline{\includegraphics[angle=-90,width=9cm,keepaspectratio,clip]{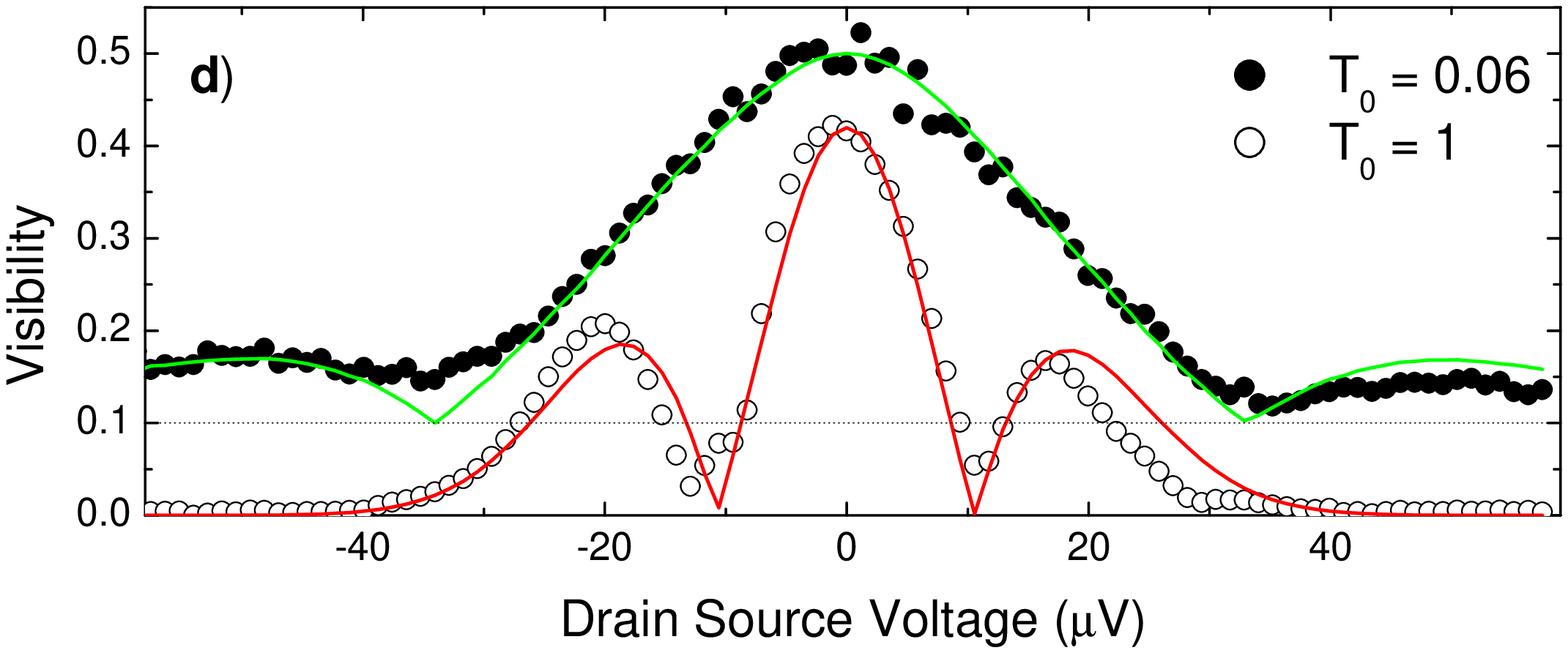}}
\caption{(Color online) \textbf{ Sample \#2 : a)} Gray plot of the
transmission $\T$ as a function of the bias voltage $V$ and the
side gate voltage $V_{SG}$. Note the $\pi$ shift of the phase when
the visibility reaches 0. \textbf{b) \& c)} $\T$ as a function of
the side gate voltage for two different values of the drain source
voltage corresponding to the dashed line of \textbf{a)} (0 and 16
$\mu$V respectively). \textbf{d)} Lobe structure of the visibility
fitted using equation (1) for a diluted and an undiluted impinging
current.}\label{Lobe2.fig}
\end{figure}

An almost perfect fit for the whole range of $\T_0$ (dilution), is
\begin{equation}
\V=\V_0e^{-V^2/2V_0^2}|1-\frac{VI_D}{V_0^2dI_D/dV}|,
\end{equation}
where $V_0$ is a fitting parameter. Equation (1) is obtained when
assuming a gaussian phase averaging  with a variance
$<\delta\varphi^2>$ proportional to $V^2$ and a length difference
$\Delta L$ small enough to neglect the energy dependence of the
phase in the observed energy range $eV\ll\hbar v_D/\Delta L$. In
such a case, the interfering part of the current $I_\sim$ is thus
proportional to $I_D\sin(\varphi)$. The gaussian distribution of
the phase leads to $I_\sim\propto
I_D\sin(<\varphi>)e^{-<\delta\varphi^2>/2}$, where $<\varphi>$ is
the mean value of the phase distribution. The measured interfering
part of the transmission, $\T_\sim = h/e^2\,dI_\sim/dV$ gives a
visibility corresponding to formula (1) when
$<\delta\varphi^2>=V^2/V_0^2$. Such behavior gives a nul
visibility accompanied  with a $\pi$ shift of the phase when
$VI_D/(V_0^2dI_D/dV)=1$. When $\T_0\sim1$, $I_D$ is proportional
to $V$ and the width of the central lobe is simply equal to
$2V_0$. However in the most general case, $dI_D/dV$ varies with
$V$. One can see in figure (\ref{Lobe.fig}) that the fit with
Equation (1) is very good, definitively showing that the existence
of one side lobe, as observed in the experiment of
ref.\cite{Neder06PRL96p016804} at $\nu=2$ (for the highest fields)
and at $\nu=1$, can be explained within our simple approach.
Concerning multiple side lobes, we cannot yet conclude if they do
arise from long range interaction as recently proposed by
ref.\cite{Sukhorukov06CM0609288}. Our geometry is different from
the one used in the earlier experiment \cite{Neder06PRL96p016804}
and the coupling between counter propagating edge states, thought
to be responsible for multiple side lobe
\cite{Sukhorukov06CM0609288}, should be less efficient here.

To check if low frequency fluctuations have an impact on the
finite bias phase averaging, we have studied another sample, with
the same geometry and fabricated simultaneously (sample \#2),
which exhibits clear interference pattern (see Figure
(4\textbf{a,b,c})). As one can remark on figure (4\textbf{d}), the
lobe structure is well fitted with our theory, definitively
showing that the gaussian phase averaging is not associated with
low frequency phase fluctuations.

\begin{figure}[h]
%06_11_24.opj[PAPER]
%06_11_28.opj[PAPER]
\centerline{\includegraphics[angle=-90,width=9cm,keepaspectratio,clip]{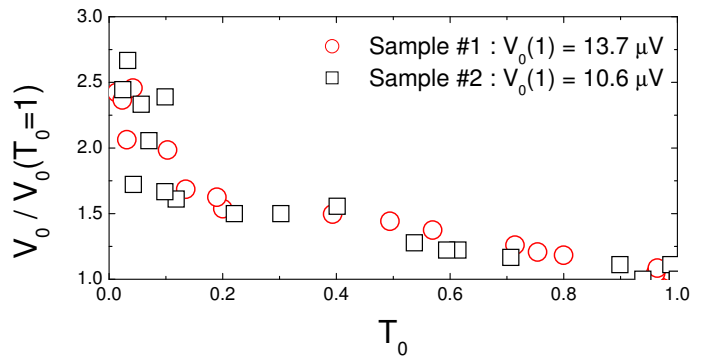}}
\caption{(Color online) $V_0$ obtained by fitting the visibility
with equation (1), normalized to $V_0$ at $\T_0=1$, as a function
of $\T_0$ at zero bias.}\label{Decay.fig}
\end{figure}

It is noteworthy that $V_0$ increases (see figure
(\ref{Decay.fig})) with the dilution, namely when the transmission
$\T_0$ at zero bias decreases. An impact of the dilution was
already observed as it suppressed multiple side lobes
\cite{Neder05CM0508024} (arXiv version of
Ref.\cite{Neder06PRL96p016804}), but the conclusion was that the
width of the central lobe was barely affected. Here, dilution
plays a clear role whose $\T_0$ dependence is the same for the two
studied samples, once normalized to the not diluted case. This
dilution effect is nevertheless not easy to explain. For example,
mechanisms like screening, intra edge scattering and fluctuations
mediated by shot noise should have maximum effect at half
transmission, in contradiction with figure (5). More generally, it
is difficult to determine if the process responsible for the phase
averaging introduced in our model is located at the beam
splitters, or is uniformly distributed along the interfering
channels. However, setting $\T_1=0.02 \ \mathrm{or} \ 0.05$,
keeping $\T_2=0.5$, leaves the lobe width unaffected. This shows
that, if located at the Quantum Point Contacts, the phase
averaging process is independent of transmission.

To summarize, we propose a statistical method to measure the
visibility of "invisible" interferences. We observe a single side
lobe structure of the visibility on stable and unstable samples
which is shown to result from a gaussian phase averaging whose
variance is proportional to $V^2$. Moreover, this variance is
shown to be reduced by diluting the impinging current. However,
the mechanism responsible for such type of phase averaging remains
yet unexplained.

The authors would like to thank M. B\"{u}ttiker for fruitful
discussions. This work was supported by the French National
Research Agency (grant n$^\circ$ 2A4002).

%\bibliography{References}

\begin{thebibliography}{10}
\bibitem{Ji03Nature422p415}
Y. Ji, Y. Chung, D. Sprinzak, M. Heiblum, D. Mahalu, and H.
Shtrikman, Nature {\bf 422}, 415 (2003).

\bibitem{Samuelsson04PRL92n026805}
P. Samuelsson, E.~V. Sukhorukov, and M. B{\"u}ttiker, Phys. Rev.
Lett. {\bf
  92},  026805  (2004).

\bibitem{Neder07CM0705.0173}
I. Neder, N. Ofek, Y. Chung, M. Heiblum, D. Mahalu, and V.
Umansky, arXiv:0705.0173 (2007).

\bibitem{Machida97SSC103p441}
T. Machida, H. Hirai, S. Komiyama, T. Osada, and Y. Shiraki, Solid
State Commun. {\bf 103}, 441 (1997).

\bibitem{Neder06PRL96p016804}
I. Neder, M. Heiblum, Y. Levinson, D. Mahalu, and V. Umansky,
Phys. Rev. Lett. {\bf 96}, 016804 (2006).

\bibitem{Litvin07PRB75n033315}
L.~V. Litvin, H.-P. Tranitz, W. Wegscheider, and C. Strunk, Phys.
Rev. B {\bf 75}, 033315 (2007).

\bibitem{Chung05PRB72p125320}
V.~S.-W. Chung, P. Samuelsson, and M. B{\"u}ttiker, Phys. Rev. B
{\bf 72},
  125320  (2005).

\bibitem{Sukhorukov06CM0609288}
E.~V. Sukhorukov and V.~V. Cheianov, Cond-Mat/0609288  .

\bibitem{Neder05CM0508024}
I. Neder, M. Heiblum, Y. Levinson, D. Mahalu, and V. Umansky,
Cond-Mat/0508024  (2005).

\end{thebibliography}

\end{document}